\documentstyle[aps,twocolumn,epsf]{revtex}
\draft
\newcommand{\ave}[1]{\left\langle #1\right\rangle}

\newcommand{\figsize}{3.0 in}

\begin{document}
\title{Turbulent self-organized criticality}
\author{Mario De Menech$^1$ and Attilio L. Stella$^{1,2}$}
\address{$^1$ INFM-Dipartimento di Fisica,\\
Universit\`a di Padova, I-35131 Padova, Italy\\
$^2$ Sezione INFN, Universit\`a di Padova, I-35131 Padova, Italy }
\date{\today} 
\maketitle

\begin{abstract}
In the prototype sandpile model of self-organized criticality
time series obtained by decomposing avalanches
into waves of toppling show
intermittent fluctuations. The $q$-th moments
of wave size differences
possess local multiscaling and global simple scaling 
regimes analogous to those holding for velocity
structure functions in fluid turbulence. The correspondence
involves identity of a basic scaling relation and of the
form of relevant probability distributions. 
The sandpile provides a qualitative analog of 
many features of turbulent phenomena.

\bgroup
\egroup
\end{abstract}


Intermittency is a characterizing feature of
turbulence~\cite{Frish,Lohse}.  Intervals of low activity and of
irregular duration are separated by sudden bursts in a wide range of
strengths. Random quantities like activity variations across fixed
space or time distances, are distributed according to non-Gaussian
laws, whose moments display also a peculiar multiscaling. At
first sight, some qualitative features of turbulent signals seem to be
recognizable in the distant context of self-organized critical
dynamics~\cite{Bak}.  In this field suitable models describe the bursts
of activity by which a system, subjected to an external inflow of energy
or matter, relaxes under the control of local, nonlinear threshold
mechanisms. The bursts are called avalanches, and their time series
show indeed some similarities to those of turbulent fluctuating
quantities. Of course, these similarities are only apparent. Indeed,
as a rule, different avalanches are just uncorrelated, and the
probability distribution functions (pdf's) of avalanche quantities do
not show multiscaling.

Quite recently it was realized
that the two-dimensional (2D) Bak-Tang-Wiesenfeld (BTW) sandpile, 
the prototype model of self organized criticality~\cite{Bak,Dhar},
obeys a peculiar form of multiscaling for the
pdf's of several avalanche measures~\cite{Tebaldi}. 
This result came after
many years of work assuming allmost unanimously
the simple finite size scaling form.
The unexpected multiscaling makes the 2D BTW one special among models
in the same class, and raises the issue of understanding its origin and
the possible analogies with other complex scaling phenomena.

In this Letter we show that one can indeed establish a surprisingly
close correspondence between BTW sandpile dynamics
and fully developed turbulence. This correspondence
includes an identity 
of the basic relations connecting the multiscaling exponents of the
relevant pdf's.

The BTW model is defined on a square lattice ($L\times L$)
box~\cite{Dhar}.  On site $i$ $z_i=0,1,2,\ldots$ is the number of
``grains''. If $z_i<4$, $\forall i$, the configuration is stable.
Grain addition to a stable configuration is made by selecting at
random a site $k$ where $z_k \to z_k+1$. If then $z_k\ge 4$, toppling
occurs, i.e. $z_k \to z_k-4$, while each nearest neighbor, $l$, of
site $k$ gets one grain ($z_l \to z_l +1$). If $k$ is at the border
grains are dissipated. The toppling of site $k$ may cause
instabilities in the neighbors, leading to further topplings at the
next microscopic time step, and so on. Thus, an avalanche made by a
total number $s\geq 0$ of topplings occurs before a new stable
configuration is reached and a new grain is added. After many
additions the system reaches a stationary critical state in which
avalanche properties are sampled.

A key notion in the approach to the BTW is that of wave decomposition
of avalanches~\cite{Ivashkevich}. The first wave is obtained as the
set of all topplings which can take place as long as the site of
addition is prevented from a possible second toppling. The second wave
is constituted by the topplings occurring after the second toppling of
the addition site takes place and before a third one is allowed, and
so on.  The total number of topplings in an avalanche is the sum of
those of all its waves.
The global sample of all waves resulting from the decomposition of a
sequence of several avalanches has exactly known scaling
properties, well confirmed numerically~\cite{Ivashkevich,Ktitarev}.
The wave size has a pdf satisfying finite size scaling $P_w(s,L)\sim
s^{-\tau_w} f_w(s/{L^{D_w}})$, where, in $d=2$, $\tau_w=1$, $D_w=2$, and
$f_w$ is a suitable scaling function. Most recently it has been shown
that the time series of successive waves sizes $\{s_m\}$, $m=1,2,..$,
has long range autocorrelation in the 2D BTW sandpile~\cite{DeMenech}.
This is not the case for other similar models, like the Manna
one~\cite{Manna}, in which the size of a wave is just uncorrelated
with that of the preceeding one~\cite{DeMenech}. The long time
autocorrelation of waves is responsible for the fact that the scaling
of the BTW avalanche size pdf, $P_{av}$, is different from that of
$P_w$~\cite{DeMenech}.  It remains a major challenge to identify
which features of the BTW wave time series are at the root
of the multifractal scaling pattern detected for
avalanches~\cite{Tebaldi}.  Here we find a key to this issue by
establishing a correspondence between $q$-th order wave size
difference moments of the sandpile, and velocity structure functions
of a turbulent fluid flow.

In fully developed turbulence one measures a velocity field $u(x)$,
with structure functions $G_q(r) =
\ave{|u(x+r)-u(x)|^q}^{1/q}$~\cite{Frish,Lohse}. $G_q$ displays a
power law $r$-dependence $G_q \sim r^{\zeta_q}$ in the inertial range
$ 1 < r/ \eta < \xi \sim (l/{\eta})\sim R^{3/4}$, where $\xi$ is a correlation
length, $R$ is the Reynolds number, and $l$ and $\eta$ are the integral 
and the dissipation scales, respectively.
The local dissipation rate $\epsilon \sim (\partial u/\partial x)^2$
has anomalous fluctuations. Indeed, the values of $\partial u/\partial
x$ have a broad pdf, with moments obeying multiscaling in $\xi$. By
matching the scaling of $|\partial u/\partial x|$ with that 
of $|u(x+\eta)-u(x)|$ in the inertial range, one also expects $G_q(r\sim
\eta)\sim \xi^{\alpha_q}$, where $\alpha_q$ is another $q$-dependent
exponent~\cite{Krug}.  On the other hand, as one considers
$u$-increments for $r/\eta \gg\xi$, the corresponding pdf's, due to the
lack of correlation, become $r$-independent and give information only
on the global sample of velocity variations over all $r$,
i.e. $G_q(r) \sim \ave{|u-\ave{u}|^q}^{1/q}$, for $r/\eta \gg \xi$. 
The pdf's of $u(x+r)-u(x)$ become also
Gaussian while multiscaling crosses over to a global
simple scaling, so that $G_q(r) \sim \xi^{\alpha}$, for $r/\eta
\gg\xi$, independent of $q$~\cite{Lohse}.

The scaling of $G_q$ is often detected by measuring the velocity field
$u$ at different times at a fixed position in space~\cite{Frish}. The
data of this time series are then converted by interpreting time
intervals as space intervals, once the average convection velocities
are known. This suggests to regard the time series $\{s_k\}$ of BTW
wave sizes in a similar perspective: also in this case we deal with a
long range correlated signal, and it is legitimate to ask whether
turbulence-like mechanisms in this signal are responsible for some
form of multiscaling.  Of course, a conversion of time into space
dependence would not make sense in this case. Thus, in the analogy $t$
will definitely replace $r$.  A structure function for the sandpile
can be defined as
\begin{equation}
F_q(t)=\ave{|s_{k+t}-s_k|^q}^{1/q},
\label{eq:defFq}
\end{equation}
In turbulence the difference in velocities is a measure of how a
perturbation of the average flow grows by propagating from one point
to another. In sandpile dynamics the difference in size between two
successive waves can also be considered as a measure of how a
perturbation evolves in time. The instantaneous entity of the
perturbation is represented by the number of topplings occurring in
the corresponding wave. In analogy with turbulence, a scaling form one
can expect for $F_q$ is~\cite{Krug}:
\begin{equation}
F_q(t)\sim {\xi}^{\alpha_q} t^{\zeta_q} f_q(t/{\xi}),
\label{eq:scalingFq}
\end{equation}
where $f_q$ are scaling functions, and $\alpha_q$ and $\zeta_q$ are
exponents possibly varying with $q$, and specified here by the same
symbols as the corresponding turbulence ones.  $\xi$ in the BTW case
is a correlation time for waves, which should scale $\sim L^{z}$.  It
plays the role of the correlation length in turbulence, and is denoted by
the same symbol here.  The long time correlation is a peculiar feature
of 2D BTW wave series, essential for establishing the correspondence
with turbulence.  Avalanche time series in general show finite correlation
time for models of self organized criticality, including the 2D BTW 
sandpile~\cite{Davidsen}.

If assumed for $G_q$, the form (2) embodies the various scaling regimes
described above for turbulence~\cite{Krug}.  Indeed, we expect $f_q(y
\to \infty)\sim y^{-\zeta_q}$ in such a way that the $t$ dependence
when $t\gg\xi$ is correctly absorbed.  In this limit one gets then
$F_q \sim \xi^{\alpha_q+\zeta_q}$.  We know further that, for
$t\gg\xi$, $F_q$ should reproduce the moment scaling of the global
wave distribution $P_w$.  Indeed, wave sizes are totally uncorrelated
for $t\gg\xi$, so that $F_q(t\gg\xi) \sim
\ave{|s-\ave{s}_{P_w}|^q}_{P_w}^{1/q}$, with $\ave{-}_{P_w} \sim
\int - P_w(s,L) ds$. Since $P_w$ has a simple, constant gap ($D_w=2$)
scaling form, $F_q(t\gg\xi) \sim \ L^2 $.  Thus, the exponents in
Eq.~(\ref{eq:scalingFq}) must satisfy the identity
\begin{equation}
\alpha_q +\zeta_q =2/z
\label{eq:alphapluszeta}
\end{equation}
for all $q$. A scaling relation analogous to 3 is expected to hold
for the corresponding turbulence exponents~\cite{Krug}.  For $1<t<\xi$
one expects $f_q \sim const.$, which leads to a power law $t$-dependence,
$F_q(t)\sim t^{\zeta_q}$, in the analog of the inertial range.
Finally, for $t \sim 1$, which corresponds to $r\sim \eta$, one
recovers the moments of the equivalent of $|\partial u/\partial x |$ in
turbulence: $F_q(1)\sim \xi^{\alpha_q}$. The interval between
successive waves, taken here as the unit of time, corresponds to
$\eta$ in the turbulent case.

\begin{figure}[tbp]
  \centerline{
  \epsfxsize=\figsize
  \epsffile{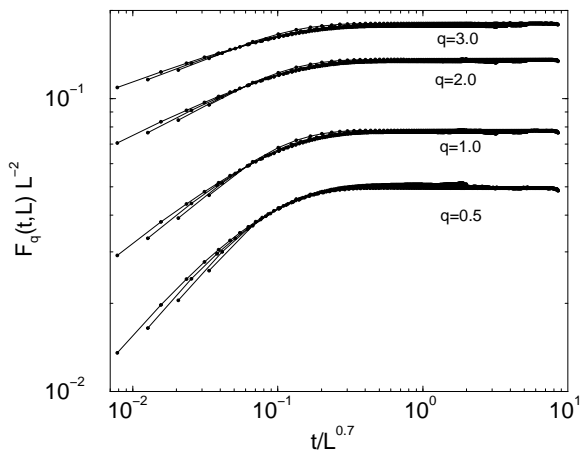}
  }
\caption{Collapse plots for $f_q$.}
\label{fig:collapse}
\end{figure}

\begin{figure}[tbp]
  \centerline{
  \epsfxsize=\figsize
  \epsffile{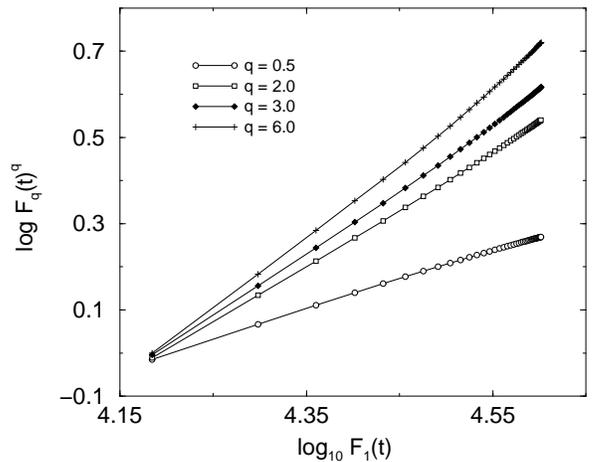}
  }
\caption{Extended mutual power law scaling between
different moments of the pdf of $|s_{k+t}-s_k|$.}
\label{fig:extended}
\end{figure}

By analyzing samples of up to $10^7$ waves for $L=128,256,512$ and
$1024$, we could verify that the scaling ansatz 2 and relation 3 are
very well satisfied.  Fig.~\ref{fig:collapse} reports collapse plots
of the quantity $y^{\zeta_q} f_q(y)$ for various $q$'s. The collapse
quality increases with $q$ and $z\sim 0.7$ is an optimal choice for
all cases.  Since $f_q$ is approximately constant for $y < 1$, from
the low $y$ part of the plots one also obtains estimates of $\zeta_q$.
The $t$-dependence of $F_q$ in the inertial range shows some degree of
extended self-similarity, a well known property of structure functions
in turbulence~\cite{Benzi}:
log-log plots of $F_q(t)^q$ vs $F_1(t)$ (Fig.~\ref{fig:extended})
are rather straight
in the range corresponding to the inertial regime. In spite of
the relatively narrow interval of variation of the moments in this range,
this allows rather precise determinations of $\zeta_q/\zeta_1$. 
These, combined with an estimate of $\zeta_1$ from the plots in
Fig.~\ref{fig:collapse} lead to the values reported
in Table~\ref{tab1}.
The exponents $\alpha_q$ reported there are determined 
by log-log plots of $F_q(1)$
versus $L$, which are definitely linear and show clear $q$-dependent
slopes.  The values of $z \zeta_q$ and $z \alpha_q$ reported in
Table~\ref{tab1} satisfy Eq.~(\ref{eq:alphapluszeta}) very well.

A key fluctuating quantity in the description of turbulent flow is the
dissipation rate averaged over spatial volumes of linear size $r$,
$\epsilon_r$~\cite{Lohse}.  In order to get further insight into the
correspondence with turbulence, we studied an analogous wave quantity,
given by the average over an interval $t$ of what should correspond to
the local dissipation rate:
\begin{equation}
\epsilon_t=\sum_{k=1}^t (s_{k+1}-s_k)^2 /t
\label{eq:epsilon}
\end{equation} 
In analogy with turbulence we expect
$\ave{\epsilon_t^q}/\ave{\epsilon_t}^q \sim t^{-\mu_q}$, in the
inertial range. Fig 3 shows plots of this quantity which clearly
manifest multiscaling and are remarkably similar to those one obtains
for $\epsilon_r$ in turbulence~\cite{Lohse}.
The plots of course do not
include the plateaus one finds in the turbulent dissipation range
($r<\eta$), which is not covered by our correspondence.  Convergence
to zero for all $q$ can also be perceived for large $t$.  These
results further support the identification of $s_{k+1}-s_k$ as the
analog of a velocity gradient in turbulence.

Thus, $F_q$ in the BTW sandpile has a scaling isomorphic to that of
$G_q$ in turbulence. So far, theoretical approaches always tried to
link avalanche to wave scaling~\cite{Priezzhev,Paczuski}. Since this
last scaling is simple and exactly known at global level, attempts
concentrated on the connection between avalanche scaling and this
regime, just overlooking the very possibility of a different local
regime and its consequences.  We find that, in the range $ 1<t<L^z$,
the wave series displays complex multiscaling structure, of the same
type as that found for velocity fields in turbulence.  This is not the
case for the Manna sandpile~\cite{Manna}, which does not show any form
of multiscaling.  In that model, due to the absence of correlation,
the global regime of the structure function is reached as soon as
$t=1$ and there is no window for local and inertial multiscaling
regimes.

The intermittent local wave regime is certainly at the basis of the
multiscaling already detected for global avalanche distributions of
the 2D BTW model~\cite{Tebaldi}. An open problem remains that of
connecting the multifractal exponents of the wave time series with
those of avalanche distributions.  Wave and avalanche time series are
separated by an infinite coarse graining in time, for $L\to \infty
$. At numerical level, one could in principle perform this coarse
graining by partitioning waves into successive sets of $n$
elements~\cite{DeMenech}. The corresponding block variables $S_{k}=
s_{n(k-1)+1}+s_{n(k-1)+2}+\ldots+s_{nk}$, $k=1,2,\ldots$ are expected
to have scaling properties approaching those of avalanche sizes in the
limit $n\to \infty$.  So, for coarse grained structure functions
$F_{q,n}= \ave{|S_{k+t}-S_k|^q}^{1/q}$, the scaling of
Eq.~(\ref{eq:scalingFq}) should cross over, for increasing $n$, to the
$t$-independent form $F_{q,n}\sim L^{\sigma_q/q}$, where $\sigma_q$ is
the multifractal moment exponent of the global avalanche size
distribution $P_{av}(s,L)$, i.e. $\ave{s^q}_{P_{av}} \sim
L^{\sigma_q}$~\cite{Tebaldi}. Unfortunately, sampling limitations make
such coarse graining extremely difficult to realize for large enough
$n$.

The analogy with turbulence goes beyond the scaling properties alone.
The pdf's of velocity differences in turbulence show an
increasing asymmetry between positive and negative increments as $r$
is reduced within the inertial range~\cite{Lohse}.  This asymmetry,
implying nonzero skewness, is an important feature of energy transfer
across different scales.  Fig.~\ref{fig:fit} reports a plot of
the pdf of $s_{k+1}-s_k$ for $L=1024$, which shows the expected slight
asymmetry.  One can also try to apply to such plots fitting
schemes used in turbulence. We adopted a scheme proposed most
recently, based on non--extensive statistics assumptions, and
extremely successfull in reproducing fluid turbulence experimental
data~\cite{Beck}. Omitting details, to be 
published elsewhere, we just 
include in Fig.~\ref{fig:fit} the obtained best fit. This is very satisfactory
within a substantial deviation range and the two parameters involved
obey within less than a percent a relation established in
Ref.~\cite{Beck}.

Summarizing, BTW wave time series produce multiscaling schemes and
pdf's isomorphic to those of fully developed turbulence. This
identifies the origin of the peculiar scaling of the 2D BTW sandpile,
compared to similar models, and gives it a novel theoretical valence.
Further elucidating why BTW waves behave like velocities
of a turbulent flow, or why $(s_{k+1}-s_k)^2$ can be viewed as a fluid
energy dissipation rate, should provide deeper insight into this
problematic model.
Multiscalings consistent with Eqs. (2-3) have been previously identified
also for a model of epitaxial growth in 1D~\cite{Krug} and, most
recently, for the directed polymer in tilted columnar disorder in
2D~\cite{Mohayaee}. In the second case careful determinations of
$\alpha_q$ and $\zeta_q$ revealed that the multiscaling is in fact
just a bi-scaling, of the kind expected for Burgers
turbulence~\cite{See}, rather than for turbulence~\cite{Frish,Lohse}.
Our results seem to definitely exclude bi-scaling in favor of genuine
turbulent multiscaling. Thus, the analogies with turbulence
discovered here include this crucial aspect.
A possible connection between turbulence and self
organized criticality was first conjectured when
this paradigm was introduced~\cite{Bak}. Our results
show that indeed there exists a precise analogy in the case
of the BTW sandpile, and that this analogy can be established 
through the
concept of waves, which are in turn peculiar to a restricted class of
models. However, this does not exclude that similar links with
turbulence could be established on a different basis~\cite{erzan} or 
also for other models, in which for example waves are 
not defined~\cite{BCT}. 

Recent studies of the statistics of solar flares have correctly
emphasized substantial differences existing between self-organized
criticality and turbulence mechanisms, for the explanation of power
law pdf's~\cite{Boffetta}. Such differences, which favor turbulence
models, can ultimately be ascribed to
the lack of long time correlations for the avalanches in the 
self organized critical models used so far to describe these phenomena.
Quite remarkably, the distinction between
the two mechanisms is however lost in the 2D BTW model, which has long
time correlated bursts and genuine turbulence, if studied at the wave
time scale. Self-organized critical models
in which activity bursts (no matter whether waves or avalanches~\cite
{Davidsen}) have
long time correlations, would certainly be better candidates to describe
flares. Of course, these correlations alone would not be necessarily 
enough to produce also turbulent scaling patterns.

Partial support from the European Network No. ERBFMRXCT980183 is
acknowledged.


\begin{figure}[tbp]
  \centerline{
  \epsfxsize=\figsize
  \epsffile{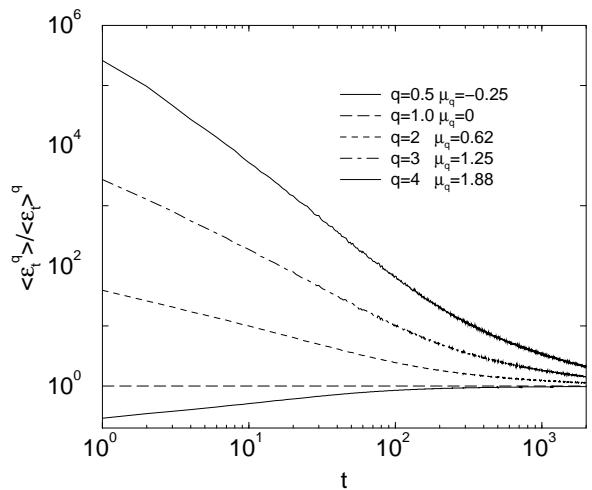}
  }
\caption{Log-log plots $\ave{\epsilon_t^q}/\ave{\epsilon_t}^q$ and
estimates of $\mu_q$.}
\label{fig:dissipation}
\end{figure}
\begin{figure}[tbp]
  \centerline{
  \epsfxsize=\figsize
  \epsffile{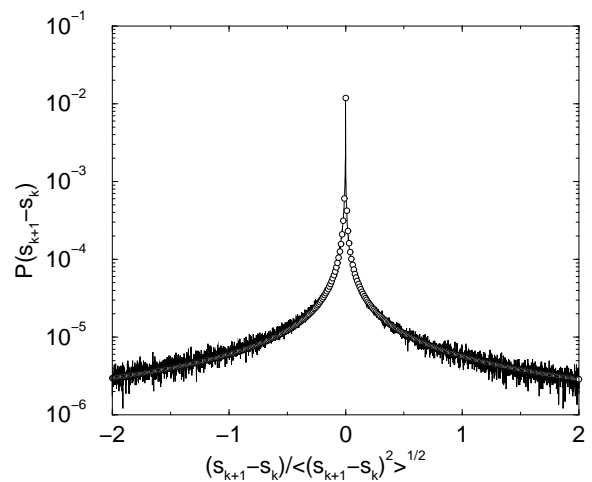}
  }
\caption{Distribution of the nomalized wave size difference
(solid line), and relative fit (empty circles).}
\label{fig:fit}
\end{figure}
\begin{table}
\begin{tabular}[htb]{|c|c|c|}
$q$ & $ z \zeta_q$ & $z \alpha_q$\\
\hline
0.5 &    0.46  &  1.43 \\      
0.8 &    0.37  &  1.58 \\ 
1.0 &    0.33  &  1.64 \\  
2.0 &    0.21  &  1.79 \\  
3.0 &    0.16  &  1.85 \\  
4.0 &    0.13  &  1.88 \\  
6.0 &    0.091 &  1.91 \\  
\end{tabular}
\caption{For each $q$ the uncertainties in both quantities 
are of the order of the discrepancy with respect to
Eq.~\ref{eq:alphapluszeta}.}
\label{tab1}
\end{table}


\begin{references}

\bibitem{Frish} U. Frisch,``Turbulence'', 
(Cambridge University Press, Cambridge, 1995).

\bibitem{Lohse} D. Lohse and S. Grossmann, Physica A {\bf 194},
519 (1993).

\bibitem{Bak}  P. Bak, C. Tang, and K. Wiesenfeld, Phys. Rev. Lett. {\bf 59},
381 (1987); Phys. Rev. A {\bf 38}, 364 (1988).

\bibitem{Dhar}  D. Dhar, Physica A {\bf 263}, 4 (1999).

\bibitem{Tebaldi}  C. Tebaldi, M. De Menech, and A.~L. Stella,
Phys. Rev. Lett. {\bf 83}, 3952 (1999);
M. De Menech, A.~L. Stella, and C. Tebaldi,
Phys. Rev. E {\bf 58}, R2677  (1998).

\bibitem{Ivashkevich} E.~V. Ivashkevich, D.~V. Ktitarev, and
V.~B. Priezzhev, Physica A {\bf 209}, 347 (1994).

\bibitem{Ktitarev} D.~V. Ktitarev, S. L\"ubeck, P. Grassberger, and
V.~B. Priezzhev, Phys. Rev. E {\bf 61}, 81 (2000).

\bibitem{DeMenech} M. De Menech and A.~L. Stella, Phys. Rev. E {\bf 62},
R4528 (2000).

\bibitem{Manna} S.~S. Manna, J. Phys. A {\bf 24}, L363 (1991).

\bibitem{Davidsen} For an exception see, e.g.,
J. Davidsen and M. Paczuski, cond-mat/0105532.

\bibitem{Krug} J. Krug, Phys. Rev. Lett. {\bf 72}, 2907 (1994).

\bibitem{Priezzhev} V.~B. Priezzhev, 
E.~V. Ivashkevich, and D.~V. Ktitarev, Phys. Rev. Lett. {\bf 76}, 2093 (1996).

\bibitem{Paczuski} M. Paczuzki and S. Boettcher, Phys. Rev. E 
{\bf 56}, R3745 (1997). 

\bibitem{Benzi} R. Benzi, S. Ciliberto, R. Tripiccione,
C. Baudet, and S. Succi, Phys. Rev. E {\bf 48}, R29 (1993).

\bibitem{Beck}  C. Beck, G.~S. Lewis, and H.~L. Swinney,
Phys. Rev. E {\bf 63}, 035303(R) (2001). For non-extensive
statistical mechanics, see C. Tsallis, Braz. J. Phys. {\bf 29}, 1 (1999).

\bibitem{Mohayaee} R. Mohayaee, A.~L. Stella, and C. Vanderzande,
Phys. Rev. Letters, {\bf 87}, 085701 (2001).

\bibitem{See} See, for example, J.~P. Bouchaud, M. Mezard, and G. Parisi,
Phys. Rev. E {\bf 52}, 3656 (1995). 
Bi-scaling would imply, e.g., a $q \zeta_q$
piece-wise linear in $q$, with just two distinct slopes.

\bibitem{erzan} A. Erzan and S. Sinha, Phys. Rev. Lett. {\bf 66},
2750 (1991).

\bibitem{BCT} The forest fire model [P. Bak, K. Chen, and C. Tang,
Phys. Lett. {\bf 147}, 297 (1990)] has been recently indicated as
a good candidate: K. Chen and P. Bak, Phys. Rev. E, {\bf 62}, 1613
(2000).

\bibitem{Boffetta} G. Boffetta, V. Carbone, P. Giuliani, P. Veltri,
and A. Vulpiani, Phys. Rev. Lett. {\bf 83}, 4662 (1999).

\end{references}
\end{document}